# Ultra-large bandwidth hollow-core guiding in all-silica Bragg fibers with nano-supports

**Guillaume Vienne**[1,*,**], **Yong Xu**[2,*], **Christian Jakobsen**[1], **Hans-Jürgen Deyerl**[3], **Jesper B. Jensen**[3], **Thorkild Sørensen**[3], **Theis P. Hansen**[1,3], **Yanyi Huang**[2], **Matthew Terrel**[2], **Reginald K. Lee**[2], **Niels A. Mortensen**[1,4], **Jes Broeng**[1], **Harald Simonsen**[1], **Anders Bjarklev**[3] **and Amnon Yariv**[2]

*(1) Crystal Fibre A/S, Blokken 84, DK-3460 Birkerød, Denmark.*
*(2) Department of Applied Physics, California Institute of Technology, Pasadena, CA 91125, USA.*
*(3) COM, Technical University of Denmark, Bldg. 345 V, DK-2800 Kgs. Lyngby, Denmark.*
*(4) Department of Micro and Nanotechnology, Technical University of Denmark, Bldg. 345 Ø, DK-2800 Kgs. Lyngby, Denmark.*

*(\*) These authors contributed equally to this work*
*(\*\*) Present address: Risø National Laboratory, Optics and Plasma Physics, PO Box 49, DK-4000 Roskilde, Denmark*
*yong@its.caltech.edu.*

**Abstract:** We demonstrate a new class of hollow-core Bragg fibers that are composed of concentric cylindrical silica rings separated by nanoscale support bridges. We theoretically predict and experimentally observe hollow-core confinement over an octave frequency range. The bandwidth of bandgap guiding in this new class of Bragg fibers exceeds that of other hollow-core fibers reported in the literature. With only three rings of silica cladding layers, these Bragg fibers achieve propagation loss of the order of 1 dB/m.

© 2004 Optical Society of America

**References and Links**
1. P. Yeh, A. Yariv, and E. Marom, "Theory of Bragg fiber," J. Opt. Soc. Am. **68**, 1196-1201 (1978).
2. J. C. Knight, J. Broeng, T. A. Birks, and P. St. J. Russell, "Photonic band gap guidance in optical fibers," Science **282**, 1476-1478 (1998).
3. R. F. Cregan et al., "Single-mode photonic band gap guidance of light in air," Science **285**, 1537-1539 (1999).
4. J. C. Knight, "Photonic crystal fibres," Nature **424**, 847-851 (2003).
5. Y. Fink et al., "Guiding optical light in air using an all-dielectric structure," J. Lightwave Technol. **17**, 2039-2041 (1999).
6. B. Temelkuran, S. D. Hart, G. Benoit, J. D. Joannopoulos, and Y. Fink, "Wavelength-scalable hollow optical fibres with large photonic bandgaps for CO2 laser transmission," Nature **420**, 650-653 (2002).
7. C. M. Smith et al., "Low-loss hollow-core silica/air photonic bandgap fibre," Nature **424**, 657-659 (2003).
8. Y. Xu, and A. Yariv, "Loss analysis of air-core photonic crystal fibers," Opt. Lett. **28**, 1885-1887 (2003).
9. F. Benabid, J. C. Knight, G. Antonopoulos, and P. St. J. Russell, "Stimulated Raman scattering in hydrogen-filled hollow-core photonic crystal fibres," Science **298**, 399-402 (2002).
10. D. G. Ouzounov et al., "Generation of megawatt optical solitons in hollow-core photonic band-gap fibers," Science **301**, 1702-1704 (2003).
11. N. A. Mortensen, and M. D. Nielsen, "Modeling of realistic cladding structures for air-core photonic bandgap fibers." Opt. Lett. **29**, 349-351 (2004).
12. T. P. White, R. C. McPhedran, L. C. Botten, G. H. Smith, and C. M. de Sterke, "Calculations of air-guided modes in photonic crystal fibers using the multipole method," Opt. Express **9**, 721-732 (2001), http://www.opticsexpress.org/abstract.cfm?URI=OPEX-9-13-721.
13. S. G. Johnson et al., "Low-loss asymptotically single-mode propagation in large-core OmniGuide fibers." Opt. Express **9**, 748-779 (2001), http://www.opticsexpress.org/abstract.cfm?URI=OPEX-9-13-748.

14. A. Argyros, "Guided modes and loss in Bragg fibres," Opt. Express **10**, 1411-1417 (2002), http://www.opticsexpress.org/abstract.cfm?URI=OPEX-10-24-1411.
15. Y. Xu, A. Yariv, J. G. Fleming, and S. Lin, "Asymptotic analysis of silicon based Bragg fibers," Opt. Express **11**, 1039-1049 (2003), http://www.opticsexpress.org/abstract.cfm?URI=OPEX-11-9-1039.
16. Y. Xu, R. K. Lee, and A. Yariv, "Asymptotic analysis of Bragg fibers," Opt. Lett. **25**, 1756-1758 (2000).
17. A. Yariv, and P. Yeh, *Optical Waves in Crystals*, (Wiley, New York, 1984).
18. J. D. Joannopoulos, R. D. Meade, and J. N. Winn, *Photonic Crystals: Molding the Flow of Light* (Princeton Univ. Press, Princeton, New Jersey, 1995).
19. E. Chow et al., "Three-dimensional control of light in a two-dimensional photonic crystal slab," Nature **407**, 983-986 (2000).
20. M. Notomi et al., "Structural tuning of guiding modes of line-defect waveguides of silicon-on-insulator photonic crystal slabs," IEEE J. Quantum Electron. **38**, 736-742 (2002).

## 1. Introduction

The concept of hollow-core Bragg fibers, in which the fiber cladding is composed of cylindrical dielectric layers with alternating refractive indices (Fig. 1(a)), was first proposed in 1978 [1]. More recently, Knight *et al.* and Cregan *et al.* demonstrated another class of hollow-core fibers, namely the photonic crystal fibers, where the cladding structure is formed by creating a two dimensional array of air holes in a high index material [2-4]. In both cases, light is confined to a central hollow-core due to Bragg reflection from the periodic cladding structure. With little overlap between the propagating modes and the cladding materials, hollow core fibers are ideal for applications involving high optical powers [5, 6]. Other potential advantages include significantly reduced nonlinearity and lower propagation loss [7, 8]. In addition, hollow core fibers also provide an attractive paradigm to study novel nonlinear optical phenomena in gas phase materials [9, 10].

One major challenge in fabricating Bragg fibers involves the identification of dielectric materials with not only a large index contrast, but also compatible rheological and thermal properties. In this paper, we report a new class of Bragg fibers that are composed of cylindrical dielectric layers in air, separated by nanoscale support bridges. Consequently, such Bragg fibers can be constructed from a single dielectric material such as a glass or a polymer. Furthermore, they have the unique property of supporting bandgap guiding over an octave frequency range, which is much larger than what has been achieved by any other hollow-core fiber [6, 7, 11-15]. Ideally, our theoretical calculations indicate that such hollow-core air-silica Bragg fibers can achieve confinement loss of the order of 0.1 dB/km with only four silica cladding rings.

This paper is organized as follows. In section 2, we theoretically analyze the properties of hollow-core air-silica Bragg fibers and demonstrate some of the unique features of this new class of Bragg fibers. In section 3, we discuss the fabrication and the experimental characterization of the hollow-core air-silica Bragg fibers in some details. The paper is summarized in section 4.

## 2. Theory and analysis

For a guided mode propagating within the hollow-core of a Bragg fiber, the temporal and the longitudinal dependence of the mode can be written as $\exp[i\omega t - i\beta z]$, where $\omega$ is the optical frequency and $\beta$ is the propagation constant. Due to the cylindrical symmetry of the fiber structure, we can further classify different guided modes according to their azimuthal quantum number $l$, with the $l$th order mode having an azimuthal dependence of $\exp[\pm il\phi]$ [1]. The polarization of a guided mode with $l=0$ is either transverse-electric (TE) or transverse-magnetic (TM), where the electric field or the magnetic field has only the azimuthal component [1, 15, 16]. The modes with $l \neq 0$ generally possess all six electromagnetic field components, and are referred as the mixed polarization (MP) mode [1, 15, 16].

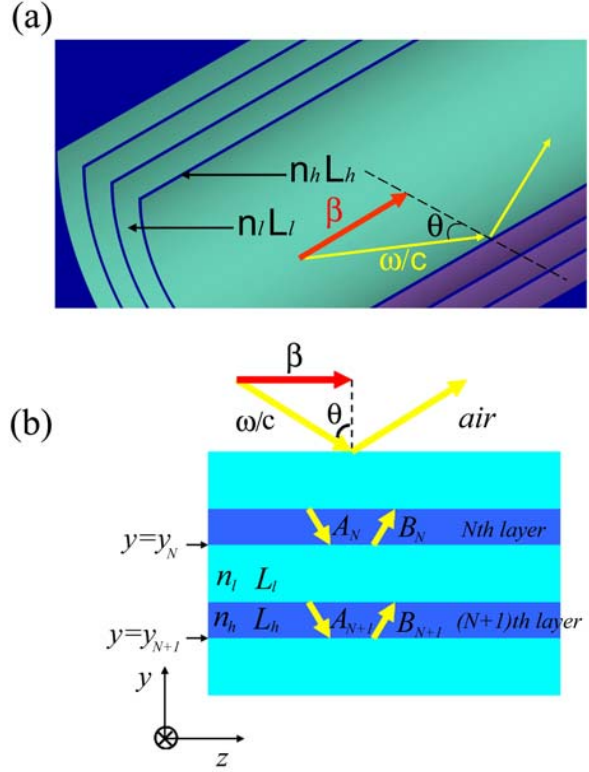

Fig. 1. (a) Schematics of a hollow-core Bragg fiber. The refractive index and the thickness of the high (low) index cladding layer are respectively $n_h$ and $L_h$ ($n_l$ and $L_l$). Photons zigzag within the hollow-core with an incident angle θ and form a propagating mode. $\omega/c$ is the vacuum wave vector. $\beta$ is the propagation constant. (b) Under the asymptotic limit, the Bragg fiber cladding layers can be well approximated by a planar Bragg stack with the same parameters. The photon wave vectors ($\omega/c$ and $\beta$) also correspond to those shown in (a).

In the asymptotic formalism in Refs. [15] and [16], the cladding field of a guided Bragg fiber mode can always be separated into a TE component and a TM component, which behave in the same way as a TE field and TM field in a planar Bragg stack. As a result, the asymptotic approach allows us to approximate a cylindrical Bragg fiber as a planar Bragg stack with the same material composition and layer thicknesses. To begin our analysis, we consider a planar Bragg stack with $N_{Bragg}$ pairs of dielectric layers, where each Bragg pair consists of a high index layer with refractive index $n_h$ and thickness $L_h$ followed by a low index layer with refractive index $n_l$ and thickness $L_l$. In the high index layer of the $N$th unit cell of the planar Bragg stack, the electric field of the TE mode takes the form of [15-17]:

$$E = A_N e^{-ik_h(y-y_N)} + B_N e^{ik_h(y-y_N)}, \quad (1)$$

where $A_N$ and $B_N$ are the amplitudes of the traveling wave within the high index layer. The transverse wave vector $k_h$ is defined as:

$$k_h = \sqrt{(n_h \omega/c)^2 - \beta^2}, \quad (2)$$

where $\beta$ is wave vector component that is parallel to the interface between Bragg stack layers, and under the asymptotic approximation, equivalent to the propagation constant of a confined Bragg fiber mode. According to Ref. [15-17], we can express the wave amplitudes in the

($N+1$)th layer ($A_{N+1}$ and $B_{N+1}$) in terms of those in the $N$th layer ($A_N$ and $B_N$) via a two by two matrix:

$$\begin{bmatrix} A_{N+1} \\ B_{N+1} \end{bmatrix} = \begin{bmatrix} a & b \\ c & d \end{bmatrix} \begin{bmatrix} A_N \\ B_N \end{bmatrix}, \qquad (3)$$

with the four coefficients $a, b, c, d$ given by [15-17]:

$$a = e^{-ik_h L_h}\left[\cos(k_l L_l) - i\frac{k_l^2 + k_h^2}{2k_l k_h}\sin(k_l L_l)\right], \qquad (4a)$$

$$b = -i\frac{k_l^2 - k_h^2}{2k_l k_h} e^{ik_h L_h} \sin(k_l L_l), \qquad (4b)$$

$$c = i\frac{k_l^2 - k_h^2}{2k_l k_h} e^{-ik_h L_h} \sin(k_l L_l), \qquad (4c)$$

$$d = e^{ik_h L_h}\left[\cos(k_l L_l) + i\frac{k_l^2 + k_h^2}{2k_l k_h}\sin(k_l L_l)\right], \qquad (4d)$$

where the parameter $k_l$ is defined as $k_l = \sqrt{(n_l \omega/c)^2 - \beta^2}$.

It is well known that the field amplitudes ($A_N$ and $B_N$) within a planar Bragg stack decay exponentially as a function of the Bragg stack unit cell number ($N$) [15, 17], which can also be inferred by repeatedly applying Eq. (3). As a result, the transmission coefficient $T$ through the planar Bragg stack also depend exponentially on the number of Bragg stack layers $N_{Bragg}$ (i.e. $T \propto D^{N_{Bragg}}$), as demonstrated in Ref. [15]. Consequently, we can define the constant $D$ as the leakage reduction factor, which characterizes how efficiently the Bragg stack reflects the incident light. Assuming an incident angle of $\theta$ (see Fig. 1), and that the ambient medium is air, we can minimize the Bragg stack transmission by choosing $L_h$ and $L_l$ such that they satisfy the quarter-wave condition [1]. Under these considerations, we can apply Eqs. (3) and (4), and use the analysis in Ref. [15], which gives the leakage reduction factor for the TE incident wave as:

$$D_{TE} = \frac{n_l^2 - \sin^2\theta}{n_h^2 - \sin^2\theta}. \qquad (5)$$

In deriving Eq. (5), the relation $\beta = (\omega/c)\sin\theta$ is used. Similarly, for the TM incident wave, we have:

$$D_{TM} = \min\left[\frac{n_l^4}{n_h^4}\frac{n_h^2 - \sin^2\theta}{n_l^2 - \sin^2\theta}, \frac{n_h^4}{n_l^4}\frac{n_l^2 - \sin^2\theta}{n_h^2 - \sin^2\theta}\right]. \qquad (6)$$

The previous analysis demonstrates that intuitively, we can regard the guided mode in a hollow-core Bragg fiber as composed of photons zigzagging within the central core and bound by the cladding Bragg stack, as illustrated in Fig. 1. At each reflection, a small amount of light passes through the Bragg cladding layers, which results in the finite loss of the Bragg fiber. Therefore, the problem of constructing a low loss Bragg fiber is equivalent to finding the planar Bragg stack with the highest possible reflection. Most hollow-core Bragg fibers in the literature have cladding structures with a large index contrast between $n_h$ and $n_l$ [5, 6, 13, 15], which according to Eqs. (5) and (6) can effectively reduce the leakage through the cladding and thus the fiber propagation loss. However, a more efficient leakage reduction can be realized if the light incidence angle $\theta$ is near 90º and the low index Bragg material is air

($n_l = 1$). From Eq. (1), it is clear that in this case the leakage reduction factors $D_{TE}$ and $D_{TM}$ are no longer limited by the index contrast of the Bragg stack, and can be arbitrarily small as $\theta$ approaches 90º. Consequentially, we expect that this new class of Bragg fibers, where both the fiber core and the low index cladding layers are composed of air, can lead to a substantially better light confinement than what can be achieved in conventional Bragg fibers [5, 6, 13-16].

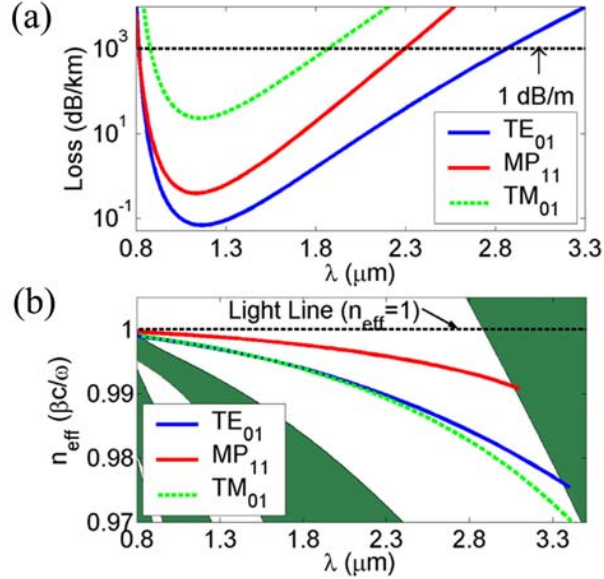

Fig. 2. In (a) and (b), we respectively give the loss and dispersion of the $TE_{01}$, $TM_{01}$, and $MP_{11}$ mode of a Bragg fiber with four cladding pairs, where we use $n_h$=1.45, $L_h$=0.37μm, $n_l$=1.0, $L_l$=4.10μm. The shaded area in (b) indicates the existence of propagating modes in the fiber cladding.

In the following, we consider a specific Bragg fiber with a hollow-core radius of 10 μm. The fiber cladding is formed by four silica layers with a refractive index of 1.45 and a thickness of 0.37 μm, separated by air layers ($n_l = 1$) with a thickness of 4.10 μm. In practice, support bridges must be introduced to separate the adjacent silica rings. However, if the support bridge thickness is much smaller than the optical wavelength of interest, we can, to a good approximation, neglect their presence and regard the region between the high index silica layers as composed entirely of air. Using an asymptotic method [15, 16], we calculate the dispersion and the loss of the fundamental TE mode ($TE_{01}$), the fundamental TM mode ($TM_{01}$), and the fundamental MP mode with $l$=1 ($MP_{11}$) in the fiber. In labeling these modes, the first subscript gives the azimuthal quantum number $l$, and the second subscript $n$ distinguishes confined modes with the same $l$ but different propagation constant. The value of $n$=1 designates the fundamental mode, which is closest to the light line. The results are shown in Fig. 2. One of the most striking features of Fig. 2(a) is that with only four silica layers, the fiber propagation loss can be reduced to less than 0.1 dB/km. Furthermore, the Bragg fiber supports low loss modes (less than 1 dB/m) in the wavelength range of 0.82 μm to 2.86 μm, almost two octaves in frequency range. To the best of our knowledge, no other hollow-core fiber [6, 7, 11-15] or photonic crystal waveguide [18-20] in the literature can support bandgap guiding over such a wide frequency range. Fig. 2(b) shows the band diagram and the dispersion of the $TE_{01}$, $MP_{11}$, and $TM_{01}$ modes for the Bragg fiber under consideration. The bandgap region extends from 0.8 μm to over 3.2 μm, a necessary condition for the ultra-large bandwidth guiding observed in Fig. 2(a). The modal dispersions shown in Fig. 2(b) are very flat, (especially that of the $MP_{11}$ mode), which makes this hollow-core Bragg fiber very attractive for applications requiring minimal signal distortion, such as the transmission of ultrafast optical pulses.

## 3. Experimental Results

To fabricate the hollow-core air-silica Bragg fibers, we use silica as the high index material, and employ a stack-and-draw method. An outer tube is filled with three tubes of decreasing sizes and a constant distance between the adjacent tubes. Within the annular gap, thin capillaries are introduced and evenly distributed. The outer, middle, and inner gaps are filled with 46, 35, and 24 capillaries, respectively. The stack of tubes and capillaries are drawn to a preform cane, which is subsequently pulled to fibers of different outer diameters (OD), of 120, 115, 110, 105, 100, 90, and 80 μm, respectively. For each fiber, the total length is more than 150m, throughout which the OD is controlled within two microns. In the following, we identify different fibers by their OD. With this fabrication procedure, the seven fibers have similar structures, except for the overall scaling factors determined by their OD.

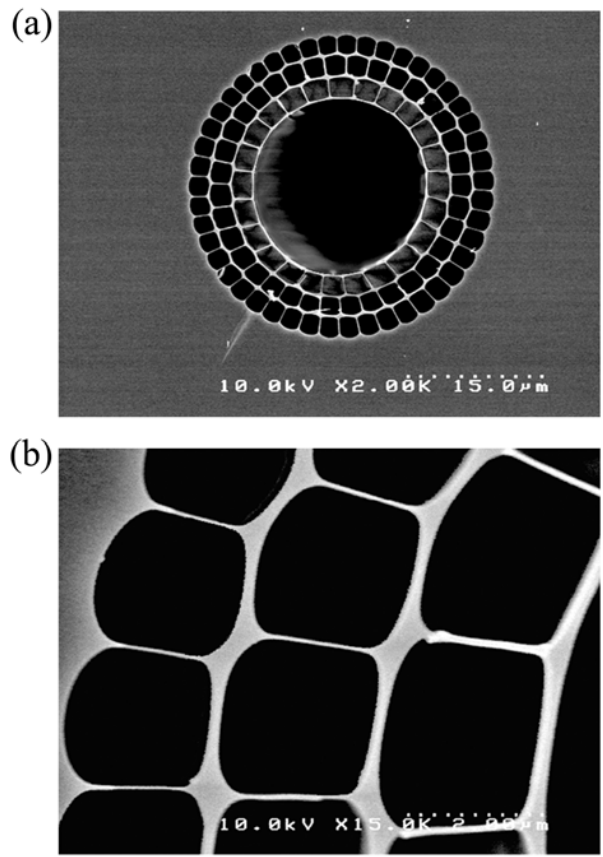

Fig. 3. The scanning electron microscope (SEM) images of the OD90 sample cross-section. In (a) and (b), we respectively show the image of the overall structure and the cladding structure.

Fig. 3 shows the scanning electron micrographs (SEMs) of the gold-coated OD90 sample facet. The three silica cladding layers can be well represented by concentric rings. The radius of the entire microstructure is approximately 17.5 μm, with a hollow core of 10 μm radius at its center. From the SEMs, the thickness of support bridges between adjacent silica layers is determined to be in the range of 60 nm to 80 nm. Assuming mass conservation throughout the fiber pulling process, we estimate the support bridge thickness to be in the range of 45 nm, which is in reasonable agreement with the results obtained from SEMs.

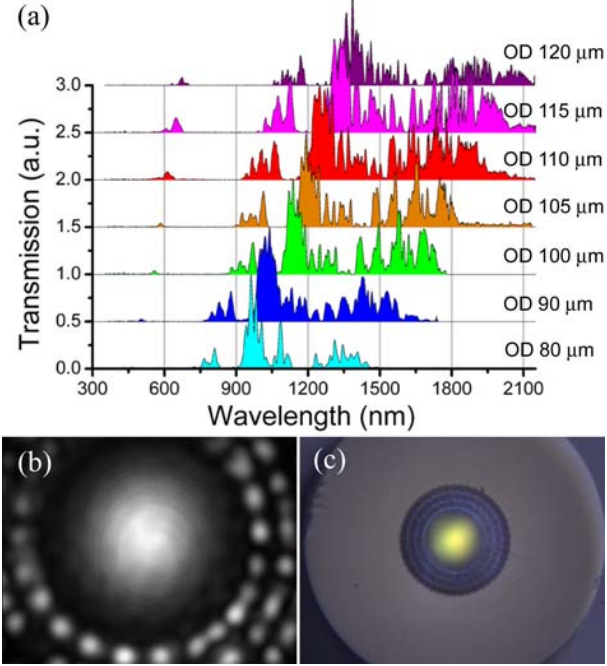

Fig. 4. (a) The experimental transmission spectra of the OD120, 115, 110, 105, 100, 90, and 80 μm samples. (b) The experimental mode profile in the OD90 sample at a wavelength of 1060 nm. (c) The image of the output facet of the OD105 sample with white light input.

We measure the Bragg fiber transmission properties by launching light from a broadband source into a large mode area (LMA) photonic crystal fiber (mode field diameter 8 μm), which is subsequently butt-coupled into the hollow-core Bragg fiber. The light transmitted through the Bragg fiber is fed into a spectrum analyzer. A Tungsten-Halogen white light source is used as the broadband source that covers the wavelength range of 400 nm to 1750 nm, and the spectra are taken with an optical spectrum analyzer (ANDO 6315E). Within the wavelength range of 1750 nm to 2150 nm, we use an ultra broadband light source (TB-1550, MenloSystems) and record the transmission spectra with a FT-IR spectrometer (Oriel MIR 8000 and DTGS detector). We choose the LMA fiber because it supports single mode propagation over the whole measurement range. In Fig. 4(a), we show the transmission spectra of seven 40 cm long fiber samples, normalized with respect to a reference measurement where only the LMA fiber is used. As expected, the fiber transmission window scales towards lower wavelength as the fiber OD decreases. Furthermore, we observe that all seven fiber samples support hollow-core guiding over approximately an octave frequency range, in accord with our previous theoretical discussions. For example, the OD90 fiber shows a broad transmission window that lies between 0.75 μm and 1.7 μm. In order to verify the existence of hollow-core confined modes, we launch light from a laser source into the fiber hollow-core and record the mode profile at the output facet with an infrared camera (Hamamatsu C2400). In Fig. 4(b), we show the experimental modal profile of the OD90 sample at a wavelength of 1060 nm. The numerous isolated small "islands" of light transmission are due to the index guiding at the intersection between the silica rings and the silica nano-supports, and are excited to provide visual indication for the position of the silica rings. The transmission spectra in Fig. 4(a) also indicate the presence of a higher order bandgap, which is manifested by the small transmission peak around 580 nm in the case of OD105 sample. The existence of a higher order bandgap can also be inferred by launching white light into the Bragg fiber core and observing the near-field images at the fiber output facet. The near-field image for the OD105 sample is shown in Fig. 4(c), which is dominated by yellow color, as one would expect from the transmission spectrum.

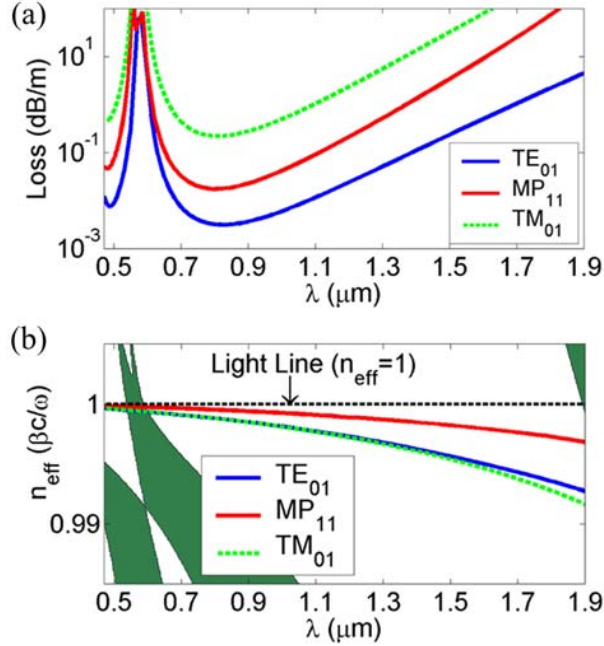

Fig. 5. The theoretical loss (a) and dispersion (b) of the $TE_{01}$, $TM_{01}$, and $MP_{11}$ mode in the OD90 sample. In calculations, we use a hollow-core radius of 10 μm (determined from SEM micrographs). The thicknesses of the inner, middle, and outer air layers are approximately 2.40, 2.27, and 2.27 μm, respectively. The thicknesses of inner, middle, and outer silica rings are 0.22, 0.28, and 0.28 μm, respectively. In calculating the cladding band diagram in (b), we assume the thickness of the silica and the air layers to be 0.28 μm and 2.27 μm, respectively. The shaded region in (b) indicates the existence of propagating modes in the fiber cladding layers.

To better understand the spectral features in Fig. 4(a), we calculate the modal loss of the OD90 fiber and compare theoretical results with experimental measurements. From SEMs, we determine that the hollow-core radius is 10 μm. The thicknesses of the inner, the middle, and the outer air layers are approximately 2.40 μm, 2.27 μm, and 2.27 μm, respectively. The dimension of the silica rings, on the other hand, can have noticeable azimuthal variation. For the innermost silica ring, where the thickness can change from 0.14 μm to 0.28 μm, we take an average value of 0.22 μm as its thickness. Similarly, we determine that the middle and the outer silica rings are approximately 0.28 μm thick. The parameters for the outer two silica/air cladding pairs are the same, and are used to obtain the band diagram for the cladding structure. The loss and the dispersion of the $TE_{01}$, $MP_{11}$, and $TM_{01}$ modes are calculated using the asymptotic approach [15, 16]. The results are shown, together with the band diagram of the cladding structure, in Fig. 5. We immediately notice that theoretical calculations predict that the fundamental transmission window covers the wavelength range from 0.6 μm up to approximately 1.8 μm, in reasonable agreement with the experimental results for the OD90 fiber (Fig. 4(a)). Furthermore, our calculations predict the second order bandgap to be around 500 nm, which also agrees well with the experimental observation.

**The propagation loss of the hollow-core Bragg fiber can be determined through a cutback method. In the measurement, we minimize the impact of index-guiding modes (as shown in Fig. 4(b)) by tuning the coupling of the input light such that the transmitted light is focused entirely within the hollow core**. From the cutback measurement of a 50-m-long fiber, we find that the loss of the OD120 sample is approximately 1.5 dB/m at a wavelength of 1.4 μm, where the transmission spectrum reaches its peak. With only three

rings of silica layers, the loss of the hollow-core Bragg fiber compares favorably with that of the Bragg fiber reported in Ref. [6], which has nine Bragg cladding pairs and has a transmission loss approximately 1 dB/m at the wavelength of 10.6 μm. We notice that the experimentally measured loss is higher than what one expects from the theoretical results in Fig. 5(a). Since the theoretical values represent the lower limit where the fiber is longitudinal uniform and cylindrically symmetric, the measured fiber loss is likely dominated by the scattering induced by the non-uniformity of the fiber cladding structures. The same non-uniformity may also explain the presence of the sharp variations in the experimental transmission spectra in Fig. 4(a) [7]. **The higher measured fiber loss can also be induced by the finite thickness of the support bridges between the adjacent silica rings, which breaks the cylindrical symmetry of the structure and cannot be taken into account by the asymptotic approach. However, we can lower the excessive loss by further reducing the thickness of the silica support bridges.**

**In this paper, the transmission window of the hollow core Bragg fiber is located within the near infrared range, with the peak transmission wavelength between 900 nm and 1550 nm. As a result, we can ignore the loss of the silica glass and assume its refractive index to be real. By increasing the overall dimension of the fiber microstructure, we can easily move the transmission window of the Bragg fiber into the middle infrared or even far infrared range. However, since the material absorption of the silica glass can be very large for the longer optical wavelength, we need to use a complex refractive index for the silica glass to calculate the fiber propagation loss. We can achieve more efficient hollow-core guiding in the longer wavelength by replacing the silica glass with a low loss polymeric material.**

## 4. Summary

In summary, we have demonstrated a new class of hollow-core Bragg fibers that requires only a single dielectric material. Such Bragg fibers have been fabricated using a stack and draw method. With only three rings of silica cladding layers, we experimentally demonstrated bandgap guiding throughout an octave frequency range, with the lowest loss at the level of 1 dB/m. Performance of this new class of hollow-core Bragg fibers can be further enhanced by decreasing the thickness of the nanoscale support bridges, improving the uniformity of the cylindrical silica rings, and adding more silica cladding layers. By scaling the dimension of the Bragg fiber appropriately and using other dielectric materials, we may achieve hollow-core guiding in other wavelength ranges of interests, such as ultraviolet, mid-infrared or even far-infrared. With the appealing properties described above, we expect that this new type of hollow-core Bragg fibers will find many applications in diverse areas such as optical communication, delivery of high power laser radiation, spectroscopy, nonlinear optics, ultrafast optics, medicine, and biology.